\documentclass{PoS}

\usepackage{psfrag}
\usepackage{epsfig}
\newcommand{\be}{\begin{eqnarray}}
\newcommand{\ee}{\end{eqnarray}}
\newcommand{\beq}{\begin{equation}}
\newcommand{\eeq}{\end{equation}}

\title{The nucleon recoil effect in antikaon-deuteron scattering at threshold}

\ShortTitle{The nucleon recoil effect in antikaon-deuteron scattering at threshold}

\author{\speaker{V. Baru},\\
        Forschungszentrum J\"ulich, Institut f\"ur Kernphysik (Theorie) and  J\"ulich Center for
Hadron Physics,  D-52425 J\"ulich, Germany and\\
          Institute for Theoretical and Experimental Physics,\\ B. Cheremushkinskaya 25, 117218 Moscow, Russia\\
                E-mail: \email{v.baru@fz-juelich.de}}
\author{E.~Epelbaum\\
       Forschungszentrum J\"ulich, Institut f\"ur Kernphysik (Theorie) and  J\"ulich Center for
Hadron Physics, D-52425 J\"ulich, Germany and\\
 Helmholtz-Institut f\"ur Strahlen- und Kernphysik (Theorie),  \\
Bethe Center for Theoretical Physics, Universit\"at Bonn, D-53115 Bonn, Germany\\
        E-mail: \email{e.epelbaum@fz-juelich.de}}
\author{A.~Rusetsky \\ 
Helmholtz-Institut f\"ur Strahlen- und Kernphysik (Theorie),  \\
Bethe Center for Theoretical Physics, Universit\"at Bonn,   D-53115 Bonn, Germany\\
E-mail:  \email{rusetsky@itkp.uni-bonn.de}}

\abstract{We investigate the role of the nucleon recoil for
antikaon-deuteron scattering using EFT. In particular, we show
that the leading correction to the static term that appears at
order  $\xi^{1/2}$ (where $\xi=m_K/M_N$) cancels completely in the
double scattering process.  We calculate the higher order recoil
corrections and study the pattern of convergence of the expansion
method. We find that the recoil correction to the double
scattering process is of order of 10-15\% of the static
contribution.}

\FullConference{International Workshop on Effective Field Theories: from the pion to the upsilon \\
		February 2-6 2009\\
		Valencia, Spain}

\begin{document}

\vspace*{-1cm}
\section{Introduction}
\label{intro}

A combined analysis of pionic hydrogen and pionic deuterium,
which has been carried out recently,
see e.g. Ref \cite{pid}, has marked a significant 
progress towards a precise 
extraction of the pion-nucleon interaction parameters at threshold. 
One may expect that, in analogy to the above case, 
 the properties of the $\bar K N$ interaction
could be investigated by using a combined analysis of kaonic hydrogen 
and kaonic deuterium. The data on this sort of bound systems are provided by
the ongoing experiment of DEAR/SIDDHARTA collaboration, which
plans to measure
the 1s energy level shift and width of kaonic hydrogen and
kaonic deuterium eventually with an accuracy of several eV \cite{Dear}.
Note however
that in contradistinction to the $\pi N$ case  the $\bar K N$
scattering length is known to be large (of order of 1 fm) and strongly
absorptive. Thus, in the kaonic case we have to solve a system of
four Deser-type equations
to determine the real and imaginary parts of the $\bar K^- p$ ($a_p$) and $\bar K^- d$
($ A_{Kd}$) scattering lengths independently:
\be
 {\Delta E_{1s}} - i \frac{ \Gamma_{1s}}{2} &=&
 -2 \alpha^3 \mu^2 { a_p} (1-2 \mu\alpha (\ln{\alpha}-1){ a_p})
\nonumber \\
 {\Delta E^d_{1s}} - i \frac{ \Gamma^d_{1s}}{2} &=&
-2 \alpha^3 \mu_d^2 {  A_{Kd}} \biggl(1-2 \mu_d\alpha
(\ln{\alpha}-1){  A_{Kd}}\biggr), \ee where $\mu (\mu_d)$
stands for the reduced mass of the $\bar K N$ ($\bar K d$) system. 
In the next step, one has to express the quantities $a_p$ and $A_{\bar Kd}$ through
the isoscalar and isovector S-wave $\bar KN$ scattering lengths $b_0,b_1$, which are defined in the
isospin-symmetric world. 
The  $K^- p$ scattering length, $a_p$, up to isospin symmetry
breaking effects (ISB) that are derived in Ref. \cite{MRR1} and
found to be significant, is just a linear combination of $b_0$ and
$b_1$: $a_p^{sym}= b_0-b_1$\footnote{Note that an extraction of
the $K^-p$ scattering length derived within chiral unitary
approaches from fits to available low-energy $K^- p$ scattering
data has been done in Ref. \cite{Borasoy}. Also a method how to extract the $\bar K N$ scattering 
lengths directly  from lattice QCD has been  proposed recently in Ref.~\cite{Lage}.}. 
The situation is totally different for the kaonic deuterium,
even in the absence of isospin breaking. During the last few decades, 
the problem of $\bar Kd$ scattering at low energy 
has been studied thoroughly within the framework of Faddeev 
equations. However,  the results of these 
 calculations are of no direct use in the analysis of the SIDDHARTA
data. In particular, these calculations do not provide
an explicit relation of $A_{\bar Kd}$ and $b_0,b_1$, which is needed for the
analysis. 
On the contrary, the multiple scattering series for the $\bar Kd$
scattering length (see, e.g.,~\cite{kamalov,MRR2})
 is well suitable for the analysis. 
It has been pointed out in Ref.~\cite{MRR2} that the non-relativistic
effective field theory (EFT) provides an ideal tool to produce the 
multiple-scattering expansion. The reason why the multiple-scattering expansion for the 
$\bar Kd$ scattering is useful, is due to the presence of two
distinct momentum scales. Whereas the $NN$ interactions and 3-body $\bar KNN$
interactions are mediated at large distances by the one-pion exchange, the
dominant long-distance contribution to the $\bar KN$ scattering comes from
the two-pion exchange. For this reason, one may treat $\bar KN$ interactions as
point-like, whereas $NN$ and $\bar KNN$ interactions will be described
by non-local potentials.
It is well known that the $\bar KN$ scattering lengths are large due to the 
presence of the subthreshold $\Lambda(1405)$ resonance, and the 
multiple-scattering series does not converge. Therefore
the full infinite set of Feynman diagrams needs to be
summed up to get a correct result for the  $\bar K d$ amplitude.
It was shown in Ref. \cite{MRR2} that the  EFT calculation designed in this manner 
at leading order 
reproduces the results of the
multiple scattering theory derived in Ref. \cite{kamalov}. Note
however that to arrive at this result the nucleon kinetic energies
have been neglected in the propagators of the Feynman diagrams,
i.e. the so-called fixed center approximation (FCA) was employed.
The validity of this approximation is the central question that 
needs to be addressed to enable a
reliable determination of the  $\bar K N$ scattering lengths from
the data. Indeed, due to a large mass of the kaon, the parameter
$m_K/M_N$ that controls the effect of the nucleon recoil is also
large ($\sim 0.5$). Thus, it is natural to conceive that the
inclusion of the nucleon recoil would affect the results of the
combined analysis of kaonic atoms. On the other hand, it is known
that the FCA results of the multiple scattering theory are in a
surprisingly good agreement with the  calculations based on the
Faddeev equations (see e.g. Ref. \cite{Gal} for a detailed
discussion of this topic) which indicates the appearance of some
sort of cancellation between different recoil contributions. In
this work we investigate the role of the nucleon recoil for  $\bar
K d$ scattering within EFT.
The key assumption for the study is that the expansion in $\xi=m_K/M_N$ 
is perturbative for each particular diagram, 
even if the multiple scattering series are not. We will verify this assumption
by  explicit calculations of the recoil effect in the double scattering process.

\vspace*{-0.5cm}
\section{What do we know about nucleon recoil?}
\label{history}

\begin{figure}
{\epsfig{file=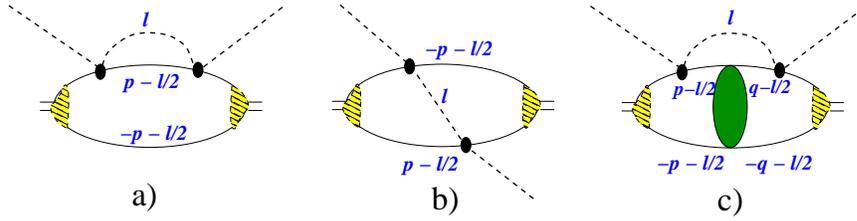, height=2.9cm}}
\caption{Diagrams of double meson scattering off nucleons in the
deuteron}
\label{double}       
\end{figure}

Historically,  the appearance of sizable cancellations in the case of
the $\pi d$ scattering length was   first pointed out in the papers
by Kolybasov et al. \cite{kolyb} and, independently, by F\"aldt
\cite{Faeldt} where it was claimed that  the naive static term provides
a good approximation for rescattering effects. Recently, the role
of the recoil effects was investigated for  $\pi d$ scattering and
for the reaction $\gamma d\to \pi^+ nn$ within EFT
\cite{Baru_rec,Lensky_gam}. It was shown, in particular, that the
3-body singularity that occurs in the $\pi NN$ intermediate state
is the origin of non-analytic corrections to the static term of
order of $\sqrt{m_{\pi}/M_N}$. It was also found that there is a
direct connection between the role of the recoil effect and the
Pauli selection rules for the intermediate NN state. In the case when
the S-wave NN interaction is forbidden by the Pauli principle, the
leading $\sqrt{m_{\pi}/M_N}$ correction cancels in the diagrams of
double $\pi N$ scattering (diagrams a) and b) in
Fig.\ref{double}, thus resulting in the small net contribution.
The situation  is different in case when the S-wave NN interaction
is allowed (see diagram c) in Fig. \ref{double}). First, the leading
$\sqrt{m_{\pi}/M_N}$  correction does not vanish here (the contributions from
diagrams a) and b) appear with the same sign). That is why it was
concluded in  \cite{Baru_rec,Lensky_gam} that the recoil effect
should be significant in this case. Secondly, the additional
diagram with the S-wave NN interaction in the intermediate state
must be also taken into account in the calculation. In the process
$\gamma d\to \pi^+ nn$, the intermediate NN interaction appears to be in the
$^1S_0$ partial wave.
Since  NN interaction in the $^1S_0$ partial
wave differs considerably to the one  in the $^3S_1$ partial wave,
 there is no {\it a priori} reason to expect any kind of
cancellation between the recoil correction from diagrams similar
to a) and b) and the contribution of diagram c).
However, for $\pi d$ and $K d$ scattering the intermediate and
final state NN interaction occurs in the same channel ($^3S_1 -
^3\!D_1$)
\footnote{Note that the relevant momenta in the integrals are
of order of $m_{\pi}$.
Therefore the contribution of the D-waves should be
significantly suppressed as compared to the S-wave part.}
.
Therefore, a combined consideration of all diagrams of
Fig.1 is needed to conclude about the recoil effect. Note that the
Pauli-allowed recoil correction for  $\pi d$ scattering goes along
with the isoscalar $\pi N$ scattering length squared which makes
this effect  negligible. This is, however, not the case for
$K d$ scattering where both the isoscalar and isovector
interactions are of a similar (and large) size.

\vspace*{-0.4cm}
\section{$\bar K d$ scattering: Recoil effect in the double scattering process}

Despite the well-known fact that
the multiple-scattering series for the $\bar K d$ scattering
are non-perturbative, at the first stage of our investigation we concentrate
on the double-scattering diagram. The aim is to demonstrate the technique
used to obtain a systematic expansion of any Feynman diagram in
(generally non-integer) powers of the parameter $\xi$. To this end,
we apply the
perturbative uniform expansion method presented in Ref.~\cite{Mohr}.  
The same results can be also obtained using the threshold 
expansion method of the Feynman diagrams in EFT developed 
in Ref.\cite{Beneke}. 
First, it can be shown that our  amplitude for
the double (see Fig.\ref{double})  as well as multiple scattering process is
expanded in  powers of $\sqrt{\xi}$ 
\be M=M_{st}+ {\sqrt{\xi}} M_1
+{\xi} M_2 + {{\xi}^{3/2}}M_3 + \cdots \label{expand}
\ee where  $M$ is related to the $\bar K d$-scattering 
length via $M=16\pi M_N(1+{\xi}/{2})A_{\bar K d}$.
The validity of this expansion and the pattern of convergence for the double 
scattering will became clear later  in this paragraph. 
To calculate the amplitudes $M_i$ we follow the
 procedure of Refs.\cite{Mohr,Beneke}. 
In particular, for double scattering one gets (see
Ref.\cite{Baru_rec} for more details) \be\nonumber
M=\!M_a\!+\!M_b\!+\!M_c
 = 2M_N (8\pi(1+\xi))^2 \left\{ {
b_0^2}\, (I_{st}\!+\!{I_0}\!+\!{I_{NN}}\!+\!\Delta I_{st})\!-\!{3 b_1^2}\,
(I_{st}\!-\!{I_1}\!+\!\Delta I_{st}) \right\}, \\
\nonumber
\hspace*{-0.3cm}{I}_{{0}({1})}\!\!=\!\! \int\! \frac{d^3p
d^3l}{(2\pi)^6}\!\! \left[\Psi^2\left(\!\vec p\!+\!\frac{\vec
l}{2}\!\right) {\pm}\Psi\left(\!\vec p\!+\!\frac{\vec
l}{2}\!\right) \Psi\left(\!\vec p\!-\!\frac{\vec
l}{2}\!\right)\right] \left[\frac{1}{\displaystyle {l^2 \!+\!
{\xi}\left(2(p^2\!+\!\gamma^2)\!+\!l^2/2\right)}}
\!-\!\frac{1}{l^2(1\!+\!\xi)}\right],
\ee
\be
\nonumber \hspace*{-0.2cm}\!\!\!{I_{NN}}\!&=&\! \frac{{\xi}}{M_N} \int\!\!\!
\frac{d^3p d^3q d^3l}{(2\pi)^9} \frac{\Psi\left(\vec
p\!+\!\frac{\vec l}{2}\right) \Psi\left(\vec q\!+\!\frac{\vec
l}{2}\right) M_{NN}(p,q,E(l))} {\left[{\displaystyle {l^2 +
{\xi}\left(2(p^2+\gamma^2)+l^2/2\right)}} \right]
\left[{\displaystyle {l^2 +
{\xi}\left(2(q^2+\gamma^2)+l^2/2\right)}} \right]},\\
\hspace*{-3.3cm} I_{st}&=& \int \frac{d^3p
d^3l}{(2\pi)^6} \frac{\Psi\left(\!\vec p\!+\!\frac{\vec
l}{2}\!\right) \Psi\left(\!\vec p\!-\!\frac{\vec
l}{2}\!\right)}{l^2}; \hspace*{0.3cm}  \ \Delta
I_{st}=-\frac{\xi}{1+\xi} I_{st} \label{ampl} \ee where $\Psi$  is
the deuteron wave function, $M_{NN}$ is the invariant NN
amplitude\footnote{On shell $M_{NN}$ is related to the scattering
amplitude $f=(k cot(\delta)-ik)^{-1}$ via $M_{NN}(k,k,E(k))=16 \pi
M_N f(k)$}, $\gamma^2=M_N \varepsilon_d$ with $\varepsilon_d$
being deuteron binding energy, and 
$E({ l})=\varepsilon_d+\frac{{ l}^2}{2m_K}+\frac{{ l}^2}{4M_N}$. 
Here, $I_{st}$ corresponds to the
LO static  (FCA) result when all $\xi$ corrections are dropped.
The recoil corrections corresponding to the Pauli-allowed
(forbidden) S-wave NN intermediate state in diagrams a) and b) of
Fig.\ref{double} are represented by the integrals $I_0$ and
$\Delta I_{st}$ ($I_1$ and $\Delta I_{st}$). 
For the Pauli-allowed NN state there is also a contribution of the diagram c)
 given by the integral $I_{NN}$.
Thus, let us define the recoil corrections as $\Delta I_1=-I_1+\Delta I_{st}$ and $\Delta I_0={I_0}+{I_{NN}}+\Delta I_{st}$
 for  isovector  and isoscalar $\bar K N$ interactions, respectively.
There are three relevant regimes in these integrals:\\
1) the regime of a 3-body singularity  (small $l$)
\begin{center}$\frac{\displaystyle l^2}{\displaystyle 2m_k} \simeq \frac{\displaystyle p^2}{\displaystyle 2M_N}
\Longrightarrow $ {$l\sim {\sqrt{\xi}}\, p, \ \ \
p\sim \langle 1/r \rangle_{\sl wf}$}; \end{center}
where $ \langle  \rangle_{\sl wf}$ denotes averaging over the deuteron wave functions.
In this regime, the integrals generate  terms with non-integer powers of $\xi$
($\sqrt{\xi}, \xi^{3/2}$,...)  in Eq.(\ref{expand}). \\
2) The  heavy-baryon regime (large $l$)
\begin{center}{$l\sim p \sim \langle 1/r \rangle_{\sl wf}$}.
\end{center}
In this regime, the integrals  produce terms with integer powers of $\xi$.\\
3) The intermediate regime, in which $\sqrt{\xi}p \ll l \ll p$.
\\
\begin{figure}[t]
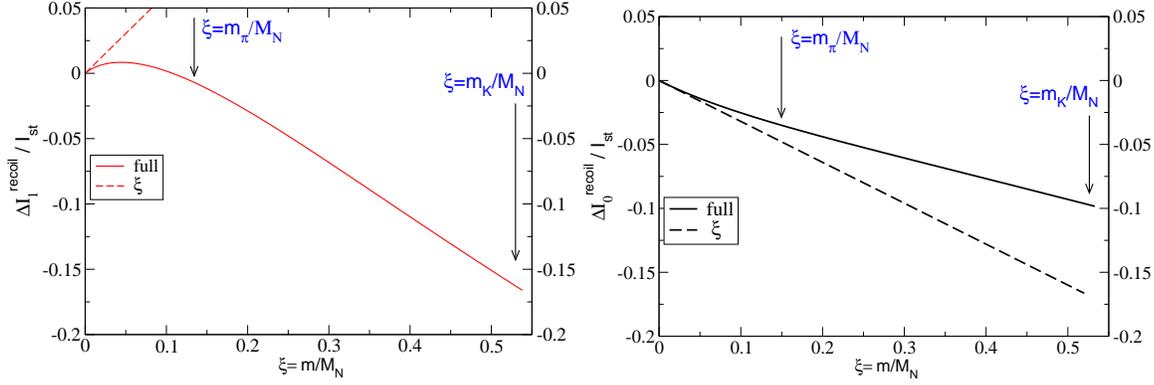

\hspace*{-0.9cm}
\vspace*{-.2cm}
\includegraphics[width=0.5\textwidth,clip=]{rec_corrI1_paper.eps}
\includegraphics[width=0.5\textwidth,clip=]{rec_corr_I0.eps}
\caption{Recoil corrections in the double scattering process for
the isovector (left panel) and isoscalar (right panel) cases.
The notation of the curves is given in the text. The arrows indicate the
results for $\pi d$- and $\bar K d$-scattering}
\label{resrec}       
\end{figure}

In the current calculation we employ  NN interaction in the
separable form which allows for an analytic study of the recoil
corrections. In this case the deuteron wave function has a Hulthen
form and its D-wave component is neglected
$$\Psi(p)=N {g(p)}({p^2+\gamma^2})^{-1}, \ \ g(p)=({p^2+\beta^2})^{-1}, \   {\rm with} \
N=\sqrt{8\pi \gamma\beta(\gamma+\beta)^3}.$$

Performing  the expansion of the integrals in each
regime  it can be easily demonstrated that there is a complete cancellation of the recoil 
corrections at order  $\sqrt{\xi}$.
One gets at this order 
\be\nonumber
\hspace*{-0.75cm}{I}_{{0}({1})}&=& -2 \xi \int \frac{d^3p
}{(2\pi)^3} \left[\Psi^2\left(\vec p\right)\right. {\pm }
\left.\Psi^2\left(\vec p\right)\right](p^2+\gamma^2) \int \frac{
d^3l}{(2\pi)^3} \frac{1}{\displaystyle  l^2}
\frac{1}{\displaystyle  {l^2 + 2{\xi}(p^2+\gamma^2))}}\\\nonumber
&=&\left\{\begin{array}{l} -\frac{\sqrt{\displaystyle
\xi}}{\displaystyle\sqrt{2}\pi}\int \frac{d^3p }{(2\pi)^3}
\Psi^2\left(\vec p\right)\sqrt{p^2+\gamma^2} {\rm\hspace*{0.65cm}
for\  isoscalar\  case}\ \ 
\nonumber\\[2mm]\nonumber\displaystyle 0  {\rm\quad\hspace*{4.85cm} for\ isovector\ case.}\ 
\\ \end{array}\right.
\ee The integral $I_{NN}$ at the same order gives \be\nonumber
\hspace*{-0.75cm}{I_{NN}}= \frac{\sqrt{\displaystyle
\xi}}{\displaystyle\sqrt{2}\pi}\int \frac{d^3p }{(2\pi)^3}
\Psi^2\left(\vec p\right)\sqrt{p^2+\gamma^2}. \ee Thus, one gets
$I_1=0$  and $I_0+I_{NN}=0$, i.e. there is no recoil corrections
at order $\sqrt{\xi}$.
Note that for $\pi d$-scattering a similar result was obtained in
Ref. \cite{Faeldt} using a potential approach.  
Thus, the   recoil corrections to the static term both for isoscalar and isovector 
$\bar K N$ interaction will start from  the term linear in $\xi$. 
 In Fig.~\ref{resrec}, we give the results for the recoil
correction (in units of the static term $I_{\sf st}$) for isovector
(left panel) and isoscalar (right panel) contributions as a
function of $\xi$. The results of our full numerical calculation  
without expanding in $\xi$ are shown by the solid curves.  
Surprisingly even for $\bar K d$ scattering the nucleon recoil effect 
turns out to be not that large as one could {\it a priori} expect. 
As can be seen from the figure the nucleon recoil for the double scattering process 
amounts just to 10-15\% of the static contribution. To understand the smallness of the effect 
more explicitly let us compare the results of the full calculation with  those of our EFT calculation. 
Our results at order $\xi$ are shown by the dashed lines.
Fig.~\ref{resrec}  shows that for the isovector case the 
full recoil correction changes its sign in the interval considered, and 
that the linear approximation in this case fails to describe the total result. 
On the other hand, the recoil correction in the isoscalar case has a constant
sign, and the linear approximation yields reasonable description.
\begin{figure}[t!]
\psfrag{rec}{$ \hspace*{-0.5cm}\displaystyle {\Delta\bf I_1}/{\bf I_{st}}$}
\parbox{7cm}{\includegraphics[scale=0.32,clip=]{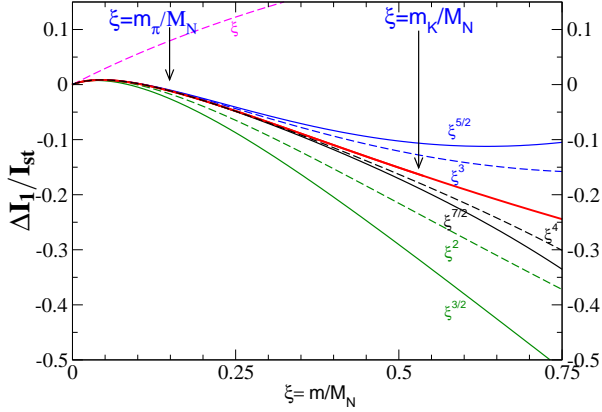}}
\hspace*{1cm}\parbox{6cm}{
\caption{\label{converg}
Recoil corrections in the double scattering process for
the isovector $\bar K N$ interaction. Convergence of the expansion in $\xi$. 
The notation of curves is explained in the text. The arrows indicate the
results for $\pi d$- and $\bar K d$-scattering. 
}}
\end{figure}
To see how does the expansion in $\xi$ converge in the case of the
isovector $\bar K N$ interaction,
we have calculated higher-order terms in $\xi$.
The results are given in Fig. \ref{converg}.  
The thick red line is again the full result, solid lines 
represent the results 
calculated up to and including some non-integer power of $\xi$,
 whereas the results shown by dashed lines include, in addition,  the next integer power of $\xi$.
Fig.~ \ref{converg} demonstrates that already at order $\xi^2$ one gets the bulk of the effect
whereas the order $\xi^4$ provides a very good approximation to the underlying result for $\bar K d$ scattering. 
It is seen that  there is a huge cancellation between the results at leading integer ($\xi$) 
and leading  non-integer ($\xi^{3/2}$) orders  that even leads to a  change of sign for the recoil effect.  
Further,  while improving convergence at smaller $\xi$, an inclusion of 
new non-integer  terms results in the oscillatory behavior around the full result at larger $\xi$. 
Thus, we conclude that the smallness of the net recoil effect is accounted for by specific 
cancellations amongst different recoil corrections.

\vspace*{-0.5cm}
\section{Summary}
\label{summary}

We have studied the nucleon recoil effect for $\bar K d$
scattering using EFT. Specifically, using the expansion method of
the Feynman diagrams in EFT, we have calculated recoil corrections
to the double scattering process in a systematic expansion in the half-integer
powers of the parameter $\xi=M_K/m_N$.
It is shown that the leading correction to the static
term, which emerges at order $\xi^{1/2}$, cancels completely both for
isoscalar and isovector types of $\bar K N$ interaction.
The origin of the cancellation for the isovector case can be
explained by the Pauli principle whereas for the isoscalar case it
stems from the orthogonality of the bound state (deuteron) and continuum
($NN$ intermediate state) wave functions in the $^3S_1$ partial
wave. The coefficients of higher order terms  in the expansion 
appear to be of a natural size and the series converges even for the value
of $\xi$ corresponding to the physical kaon mass. 
The relative smallness of the net recoil effect can be explained by specific cancellations 
of recoil corrections at different orders. 
 A more detailed discussion of this and other aspects can be found in Ref.~\cite{we}.

\vspace*{-0.5cm}

\acknowledgments

\vspace*{-0.3cm}

We would like to thank U.-G.~Mei{\ss}ner,  A. Gal and A. Kudryavtsev for interesting discussions. 
Work was supported in parts by funds provided from the Helmholtz
Association (grants VH-NG-222, VH-VI-231) and by the DFG (SFB/TR 16
and DFG-RFBR grant 436 RUS 113/991/0-1) and the EU HadronPhysics2
project (Grant agreement n. 227431).  V. B. acknowledges the support of the
Federal Agency of Atomic Research of the Russian Federation.
A.R. acknowledges financial support 
of the Georgia National Science Foundation (Grant \# GNSF/ST08/4-401).


\vspace*{-0.4cm}


\begin{thebibliography}{}

%
\bibitem{pid}
S.~R.~Beane et .,
  Nucl.\ Phys.\  A {\bf 720}, 399 (2003) [hep-ph/0206219];
U.-G.~Mei\ss ner et al.,
  Phys.\ Lett.\  B {\bf 639}, 478 (2006)   [nucl-th/0512035];
 V.~Lensky et al.,
  Phys.\ Lett.\  B {\bf 648}, 46 (2007),  [nucl-th/0608042];
  V.~Baru et al.,   
  Phys.\ Lett.\  B {\bf 659}, 184 (2008)   [arXiv:0706.4023 [nucl-th]];
V.~Baru et al.,
 In the Proceedings of MENU 2007, 
 Julich, Germany,  Sep 2007,  127
  [arXiv:0711.2743 [nucl-th]];
  J.~Gasser et al.,
  Phys.\ Rept.\  {\bf 456},  167 (2008), [arXiv:0711.3522 [hep-ph]].

\bibitem{Dear}
 G.~Beer {\it et al.}  [DEAR Collaboration],
  Phys.\ Rev.\ Lett.\  {\bf 94}, 212302 (2005);  C. Curceanu,  Proceedings of EXA 2008, Vienna, Austria,
Sep 2008,  Hyperfine Interactions,  DOI: 10.1007/s10751-009-0071-0   

\bibitem{MRR1}
  U.-G.~Mei{\ss}ner, U.~Raha and A.~Rusetsky,
  Eur.\ Phys.\ J.\ C {\bf 35}  349 (2004)
  [hep-ph/0402261];  R.~H.~Dalitz and S.~F.~Tuan, Ann. Phys. {\bf 8} 100, (1959).

\bibitem{Borasoy}
  B.~Borasoy, U.~-G.~Meissner and R.~Nissler,
  PRC {\bf 74}, 055201 (2006) [hep-ph/0606108].

\bibitem{Lage}
  M.~Lage, U.~G.~Meissner and A.~Rusetsky,
  arXiv:0905.0069 [hep-lat].




\bibitem{kamalov}
  S.~S.~Kamalov, E.~Oset and A.~Ramos,
  Nucl. Phys. A {\bf 690}, 494 (2001)
  [nucl-th/0010054];  R.~Chand and R.~H.~Dalitz,  
      Annals Phys.\  {\bf 20} (1962) 1.
\bibitem{MRR2}
  U.-G.~Mei{\ss}ner, U.~Raha and A.~Rusetsky,
  Eur.\ Phys.\ J.\  C {\bf 47}, 473 (2006)
  [nucl-th/0603029]; A.~Rusetsky, Proceedings of MENU 2007, 
 Julich, Germany,  Sep 2007,  162


\bibitem{Gal}
  A.~Gal,
  Int.\ J.\ Mod.\ Phys.\  A {\bf 22}, 226 (2007)
  [nucl-th/0607067].


\bibitem{kolyb}
V.~M.~Kolybasov and V.~G.~Ksenzov,
Zh.\ Eksp.\ Teor.\ Fiz.\  {\bf 71}  13 (1976)

\bibitem{Faeldt}
G.~F\"aldt,
Phys. \ Scripta {\bf 16}, 81 (1977)

\bibitem{Baru_rec}
  V.~Baru et al.,
  Phys.\ Lett.\  B {\bf 589}, 118 (2004)
  [nucl-th/0402027].

\bibitem{Lensky_gam}
  V.~Lensky et al.,
  Eur.\ Phys.\ J.\  A {\bf 26}, 107 (2005)
  [nucl-th/0505039].


\bibitem{Mohr}  R.~F.~Mohr et al.,
  Annals Phys.\  {\bf 321}, 225 (2006)  [nucl-th/0509076]; R.~F.~J.~Mohr,
  PhD thesis, [nucl-th/0306086]


\bibitem{Beneke}
  M.~Beneke and V.~A.~Smirnov,
  Nucl.\ Phys.\  B {\bf 522}, 321 (1998)
  [hep-ph/9711391].


\bibitem{we} V.~Baru, E.~Epelbaum and A.~Rusetsky,
 Eur. Phys. J. A, {\bf 42}, 111   [arXiv:0905.4249 [nucl-th]]



\end{thebibliography}
\end{document}